\documentclass{article}%
\usepackage{amsfonts}
\usepackage{amsmath}
\usepackage{amssymb}
\usepackage{graphicx}%
\newtheorem{theorem}{Theorem}

\newtheorem{remark}[theorem]{Remark}

\begin{document}

\title{Periodic first integrals for Hamiltonian systems of Lie type.}
\author{Ruben Flores Espinoza\\Universidad de Sonora, Hermosillo, Mexico}
\maketitle

\begin{abstract}
We prove the existence of a Lie algebra of first integrals for time dependent
Hamiltonian systems of Lie type. Moreover, applying the Floquet theory for
periodic Euler systems on Lie algebras, we show the existence of an abelian
Lie algebra of periodic first integrals for periodic Hamiltonian systems. An
application to the dynamics of a nonlinear oscillator is given.

\end{abstract}

\section{\bigskip Introduction.}

The existence of first integrals for time dependent Hamiltonian systems is a
very important topic in the theory of differential equations and its
applications. In general, for such class of systems, there is no time
independent constants of motion. This is true even for 1-dimensional classical
Hamiltonian systems of the form%
\[
H(q,p,t)=\frac{1}{2}p^{2}+V(t,q)
\]
corresponding to the motion of a particle under a time dependent potential
$V(t,q).$

In this article, we deal with the class of time dependent Hamiltonian systems
which can be represented as a linear combination of Hamiltonian systems
closing under the Lie bracket in a finite dimensional Lie algebra and having
as coefficients scalar functions of time. These systems has been studied by S.
Lie and can be considered as a generalization of linear systems but with a
nonlinear superposition rule \cite{[L]}. Recently the study of Lie systems has
revived and new approaches and applications has been done by Ibrahimov
\cite{[Ib]}, Winternitz and coworkers \cite{[W]} and Cari\~{n}ena and
coworkers \cite{[CLR1]}\cite{[CLR2]}.

In this article, we study the existence of first integrals for Lie systems
generated by Hamiltonian vector fields closing in a finite Lie algebra. Under
the assumption of the existence of Hamiltonian functions closing under the
Poisson bracket in a Lie algebra isomorphic to that generated by the original
vector fields,\ we prove the existence of a Lie algebra of first integrals for
time dependent Hamiltonian systems of Lie type. Moreover, applying the Floquet
theory for periodic Euler systems on Lie algebras, we prove for $T-$periodic
Hamiltonian systems of Lie type, the existence of an abelian Poisson algebra
of $2T-$periodic first integrals and we also give conditions for the existence
of a Lie algebra of $T-periodic$ first integrals.

Finally, we include an application to the Milne-Pinney system that describes
the time evolution of a an isotonic oscillator \cite{[M]}, \cite{[P]} founding
for that case a periodic first integral similar to that given by Lewis for
time depending oscillators \cite{[Lw]}. \ \ 

\bigskip

\section{Time dependent Hamiltonian systems of Lie type.}

Let be $(P,\omega)$ a symplectic manifold. A time dependent Hamiltonian vector
field $X$ on $P$ is called a \textit{Hamiltonian vector field of Lie type} if
$X$ can be written in the form%
\begin{equation}
X(t,x)=%
{\displaystyle\sum_{i=1}^{n}}
b_{i}(t)X_{i}(x),\text{ \ }t\in%
\mathbb{R}
,\text{ }x\in M \label{HLS}%
\end{equation}
where $\left\{  b_{i}(t)\mid i=1,...,n\right\}  $ are smooth real functions
and the vector fields $\left\{  X_{i}\mid i=1,...,n\right\}  $ are Hamiltonian
vector fields which close under the Lie bracket on a n-dimensional real Lie
algebra $\mathfrak{g}$ of vector fields, i,e. there exists $n^{3\text{ }}$real
numbers $\lambda_{ij}^{k}$ such that
\[
\lbrack X_{i,}X_{j}]=\sum_{k=1}^{n}\lambda_{ij}^{k}X_{k},\text{ }\forall
i,j=1,...,n.
\]
If the scalar functions $b_{i}(t)$ are $T-periodic,$ $b_{i}(t+T)=b_{i}(t),$
$i=1,...,n,$ the Lie system is called a $T-$\textit{periodic Hamiltonian
system of Lie type.}

Let us take Hamiltonian functions $H_{i}(x),$ $i=1,...,n$ associated to the
basis for $\mathfrak{g}$ given by the vector fields $X_{i},i=1,...,n,$ and
suppose we can choose Hamiltonian functions $H_{X}$ for the Hamiltonian vector
fields $X$ of $\mathfrak{g}$ in such way that
\begin{align}
H_{\alpha X+\beta Y}  &  =\alpha H_{X}+\beta H_{Y},\text{ \ \ }\alpha,\beta\in%
\mathbb{R}%
\label{Ass1}\\
\left\{  H_{X},H_{Y}\right\}   &  =H_{[X,Y]},\text{ \ }X,Y\in\mathfrak{g,}
\label{Ass2}%
\end{align}
where $\left\{  H_{X},H_{Y}\right\}  $denotes the Poisson bracket generated by
the symplectic structure $\omega.$In usual terms the above assumption means
the existence of a homomorphism $X\rightarrow H_{X}$ between $\mathfrak{g}$
and the Lie algebra of functions on $P.$

A non-locally constant smooth function $I:%
\mathbb{R}
\times P\rightarrow%
\mathbb{R}
$ is called a \textit{first integral} for (\ref{HLS}) if it takes constant
values on the integral curves of (\ref{HLS}). If $I(t,x)$ is a periodic
function$\ $on $t$ will be called a \textit{periodic first integral}. In terms
of the Poisson brackets on $M$, any first integral $I$ satisfies
\[
\frac{\partial I}{\partial t}+\sum_{i=1}^{n}b_{i}(t)\left\{  H_{i},I\right\}
=0.
\]

\bigskip To begin, we propose an ansatz for first integrals in the form%
\begin{equation}
I(t,x)=\sum_{k=1}^{n}p_{k}(t)H_{k}(x) \label{FirstInt2}%
\end{equation}
for smooth functions $p_{k}(t),$ $k=1,...,n$. In this case, functions
$p_{i}(t)$ have to satisfy the linear system%
\begin{equation}
\frac{\partial p_{k}}{\partial t}+\sum_{i=1}^{n}\sum_{j=1}^{n}\lambda_{ij}%
^{k}b_{i}(t)p_{j}(t)=0,\text{ \ }k=1,...,n. \label{Eulsyst1}%
\end{equation}
If we denote by $\mathcal{B=}\left\{  e_{1},...,e_{n}\right\}  $ the basis for
$\mathfrak{g}$ corresponding to the vector fields $\left\{  X_{i}\mid
i=1,...,n\right\}  ,$ the system $($\ref{Eulsyst1}) becomes the Euler system
on $\mathfrak{g}$%
\begin{equation}
\frac{d\xi}{dt}=-[\phi(t),\xi],\text{ \ \ }\xi=\sum_{k=1}^{n}p_{k}e_{k}
\label{Eulsyst2}%
\end{equation}
where $\phi(t)=$ $\sum_{i=1}^{n}b_{i}(t)e_{i}$ is a smooth curve on
$\mathfrak{g.}$

Therefore, each solution of (\ref{Eulsyst2}) gives us in (\ref{FirstInt2}) a
first integral for (\ref{HLS}). If we denote by $F(t)=\left(  \left(
F^{ij}(t)\right)  \right)  $ the fundamental matrix of the linear system
(\ref{Eulsyst2}) with $F(0)=I$, any solution of (\ref{Eulsyst2}) has the form
$F(t)\alpha$ with $\alpha\in\mathfrak{g}$ and we have a basis of solutions
given by%
\[
F(t)e_{i}=\sum_{j=1}^{n}F^{ij}(t)e_{j}.\text{ \ }i=1,...,n
\]
Moreover, if $\alpha=\sum_{j=1}^{n}\alpha_{j}e_{j}\in\mathfrak{g,}$ the family
of first integrals%
\begin{equation}
I_{\alpha}(t,x)=\sum_{j=1}^{n}\sum_{k=1}^{n}\alpha_{k}F^{kj}(t)H_{j}(x),\text{
\ }\alpha\in\mathfrak{g} \label{FirstIntegral2}%
\end{equation}
generate a Lie algebra isomorphic to $\mathfrak{g}$%
\[
\left\{  I_{\alpha},I_{\beta}\right\}  =I_{[\alpha,\beta]},\text{ \ }%
\alpha,\beta\in\mathfrak{g}%
\]
Note that each element $\beta$ of the center $\mathfrak{z}$ of $\mathfrak{g}$
is a singular point of the Euler system (\ref{Eulsyst2}) and gives place to
the time independen first integral%
\[
I_{\beta}(x)=\sum_{j=1}^{n}\beta_{_{j}}H_{j}(x),\text{ \ }\beta=\sum_{k=1}%
^{n}\beta_{k}e_{k}\in\mathfrak{z.}%
\]
The space of first integrals
\[
\left\{  I_{\beta}\text{ where }\beta\in\mathfrak{z}\right\}
\]
is an abelian Lie algebra of first integrals of (\ref{HLS})$.$

\begin{theorem}
The time dependent Hamiltonian system of Lie type (\ref{HLS}), under the
assumptions (\ref{Ass1}),(\ref{Ass2}), possesses a Lie algebra of first
integrals isomorphic to its associated the Lie algebra $\mathfrak{g.}$ If the
center $\mathfrak{z}$ of $\mathfrak{g}$ is non trivial, the system has an
abelian Lie algebra of time independent first integrals isomorphic to the
center $\mathfrak{z.}$
\end{theorem}

\bigskip

\begin{remark}
If the Lie algebra $\mathfrak{g}$ associated to the Hamiltonian system of Lie
type (\ref{HLS}) has no trivial center, we can take for the quotient Lie
algebra $\mathfrak{g}/\mathfrak{z}$ a basis $\mathcal{D=}\left\{  \left[
K_{1}\right]  ,\left[  K_{2}\right]  ,...,\left[  K_{k}\right]  \right\}  $
where $k=dim(\mathfrak{g}/\mathfrak{z)}$ and $K_{i}=K_{i}(x)$ $i=1,...,k$ are
some fixed Hamiltonian functions representants of its class. Then each of the
initial Hamiltonian vector fierlds $X_{j}(x),$ $j=1,..,n$ can be written in
the form $X_{j}=%
{\displaystyle\sum_{i=1}^{k}}
c_{ji}Y_{i}(x)+Z_{j}(x),$ $j=1,...,n$ where $c_{ij}$ are scalars, $Y_{i}(x)$
for $i=1,...,k$ is the Hamiltonian vector fields with Hamiltonian function
$K_{i}$ and $Z_{i}(x)$ are Hamiltonian vector fields belonging to the center
$\mathfrak{z.}$ To the initial time dependent Hamiltonia vector field
(\ref{HLS}) we can associate the Hamiltonian vector field of Lie type%
\[
Y=%
{\displaystyle\sum_{i=1}^{k}}
{\displaystyle\sum_{j=1}^{n}}
b_{j}(t)c_{ji}Y_{i}(x)
\]
generated by the Hamiltonian vector fields $Y_{i}(x)$ which close under the
Lie bracket into the Lie algebra $\mathfrak{g}/\mathfrak{z.}$ Then we have
$X=Y+Z$ with $[Y,Z]=0$ and each first integral $\ J_{\mu}(t,x)$ of $Y$ having
the form
\[
J_{\mu}(t,x)=\sum_{s=1}^{k}q_{s}(t)K_{s}(x)
\]
is also a first integral for (\ref{HLS}). The previous discussion allows us to
reduce the search of first integrals of the form (\ref{FirstInt2}) to the case
of Lie algebras with trivial center.
\end{remark}

\section{\bigskip Floquet theory for periodic Euler systems.}

The Euler system (\ref{Eulsyst2}) with $\phi(t)$ a $T-$periodic curve on
$\mathfrak{g,}$ is a periodic linear system on $\mathfrak{g.}$ The fundamental
matrix $F(t)$ with $F(0)=I$ preserves the Lie algebra structure%
\begin{equation}
\lbrack F(t)x,F(t)y]=F(t)[x,y],\text{ \ }\forall x,y\in\mathfrak{g}
\label{AutProp}%
\end{equation}
and
\[
F(t+T)=F(t)\circ M
\]
where $M=F(T)$ is the monodromy matrix. To the Euler system on $\mathfrak{g}$
(\ref{Eulsyst2}) we associate the $T-$periodic Lie system%
\begin{equation}
\frac{dA}{dt}=-ad_{\phi(t)}\circ A,\text{ \ }A\in Ad(\mathfrak{g)}
\label{LSADgrop}%
\end{equation}
on the adjoint group $Ad(\mathfrak{g)}$ generated by the linear operators
$e^{ad_{\alpha}}$ where $\alpha\in\mathfrak{g}$ and $ad_{\alpha}%
(\beta)=[\alpha,\beta],$ $\ \beta\in\mathfrak{g.}$ The Lie system
(\ref{LSADgrop}) possesses as fundamental solution the curve $F(t)\in$
$Ad(\mathfrak{g).}$If the center $\mathfrak{z}$ of the algebra is non trivial,
each of its elements are proper vectors of each element of the matrix Lie
group $Ad(\mathfrak{g)}$ with proper value 1. In this case, for each
$\alpha\in\mathfrak{z\,\ }$the curve $F(t)\alpha$ is a periodic solution of
(\ref{Eulsyst2}) and we also have in (\ref{FirstIntegral2}) a periodic first
integral for (\ref{HLS}).

To prove the existence of periodic first integrals for (\ref{HLS}) when
$\mathfrak{g}$ has trivial center, we consider $Ad-$invariant symmetric
bilinear form on $\mathfrak{g}$ given by the Killing form%
\begin{align*}
\left\langle ,\right\rangle _{K}  &  :\mathfrak{g\times g\rightarrow%
\mathbb{R}
}\\
\left\langle x,y\right\rangle _{K}  &  =-tr(ad_{x}\circ ad_{y}),\text{
\ }x,y\in\mathfrak{g}%
\end{align*}
The Ad-invariance of $\left\langle ,\right\rangle _{K}$ takes the form
\begin{equation}
\left\langle \lbrack\alpha,\beta],\gamma\right\rangle _{K}+\left\langle
\beta,[\alpha,\gamma]\right\rangle _{K}=0,\text{ }\forall\alpha,\beta
,\gamma\in\mathfrak{g} \label{K2}%
\end{equation}
From (\ref{K2}), the fundamental matrix $F(t)$ of (\ref{Eulsyst2}) preserves
the bilinear form
\[
\left\langle F(t)\alpha,F(t)\beta\right\rangle _{K}=\left\langle \alpha
,\beta\right\rangle _{K},\text{ }\forall\alpha,\beta\in\mathfrak{g}%
\]
and particularly its monodromy operator satisfies
\[
\left\langle M\alpha,M\beta\right\rangle _{K}=\left\langle \alpha
,\beta\right\rangle _{K},\text{ }\forall\alpha,\beta\in\mathfrak{g.}%
\]

Consider the complexification of the algebra $\mathfrak{g}^{%
\mathbb{C}
}$ and the natural extension of the monodromy operator $M$ to $\mathfrak{g}^{%
\mathbb{C}
}.$If $\lambda$ is a proper value of $M$ with proper vector $\alpha
\in\mathfrak{g}^{%
\mathbb{C}
},$ we have%
\[
\left\langle M\alpha,M\bar{\alpha}\right\rangle _{K}=\lambda\bar{\lambda
}\left\langle \alpha,\bar{\alpha}\right\rangle _{K}=\left\langle \alpha
,\bar{\alpha}\right\rangle _{K}%
\]
and $\lambda=e^{i\theta\text{ }}$ for some $\theta\in%
\mathbb{R}
$ if $\left\langle \alpha,\bar{\alpha}\right\rangle _{K}\neq0.$ A proper
vector $\alpha$ of $M$ will be called $\ $\textit{admissible} if $\left\langle
\alpha,\bar{\alpha}\right\rangle \neq0.$ The possible proper values associated
to admissible proper vectors are $\pm1$ or $e^{i\theta\text{ }}$ with
$\theta\neq2\pi k,$ $\forall k\in%
\mathbb{Z}
.$ If $\lambda=1,$ the solution $F(t)(\frac{\alpha+\bar{\alpha}}{2})$ is a
periodic real solution; if $\lambda=-1,$ the solution $F(t)(\frac{\alpha
+\bar{\alpha}}{2})$ is anti-periodic real solution $F(t+T)\alpha=-F(t)\alpha
.$If $\lambda=e^{i\theta\text{ }}$ with $\theta\neq2\pi k,$ $\forall k\in%
\mathbb{Z}
,$ we can take the real vector%
\[
\delta=\frac{1}{2i}[\alpha,\bar{\alpha}]
\]
and check using (\ref{AutProp}) that $M\delta=\delta,$ and we have taht
$F(t)\delta$ is a $T$-periodic real solution of (\ref{Eulsyst2}).

One can single out the following cases in which $\mathfrak{g}$ admitts a
Ad-invariant bilinear and non-degenerate forms and consequently all proper
vectors of $M$ are admissible. (See \cite{[D-K]})

a) The Lie algebra is semisimple,%
\[
\mathfrak{g=[g,g]}%
\]
and $\left\langle ,\right\rangle $ is defined as the Killing form%
\[
\left\langle x,y\right\rangle _{\mathfrak{g}}=-tr(ad_{x}\circ ad_{y}),\text{
\ }x,y\in\mathfrak{g}%
\]

b) The Lie algebra is compact, ie. the Killing form is negative semi-definite
and its kernel is equal to the center of $\mathfrak{g,}$then there
exists$\mathfrak{\ }$an Ad-invariant inner product on $\mathfrak{g.}$

\bigskip We summarize the above discussion with the following

\begin{theorem}
The $T-$periodic Hamiltonian system (\ref{HLS}) possesses an abelian Poisson
algebra of $2T-$periodic first integrals. Moreover, if the Lie algebra
$\mathfrak{g}$ admits an $Ad-$invariant bilinear form, the system (\ref{HLS})
has an abelian Poisson algebra of $T-$periodic first integrals.
\end{theorem}

\section{\bigskip Application: The periodic Milne-Pinney equation.}

The second order nonlinear differential equation
\begin{equation}
\frac{d^{2}y}{dt^{2}}+w(t)^{2}\frac{dy}{dt}+cy^{-3}=0 \label{M-PEq.}%
\end{equation}
describes the time evolution of an oscillator with inverse quadratic potential
and shares with the harmonic one the property of having a period independent
of the energy \cite{[CH-V]}. The equation (\ref{M-PEq.}) obeys a superposition
rule and first integrals have been obtained in \cite{[C-L-R]}. Here, we apply
our previous results and show \ the existence of a $sp(1,R)$ algebra of first
integrals for equation (\ref{M-PEq.}). Moreover, we consider the periodic case
in give a periodic first integral in terms of the periodic solution of the a
periodic Euler equation on $sp(1,R)$.

Consider the symplectic manifold $(M,\Omega),$ where $M=T^{\ast}%
\mathbb{R}
^{+}$with global coordinates $(q,p),$ $q>0$ and symplectic form $\Omega
=dq\wedge dp.$ The periodic Milne-Pinney system is given by%
\begin{align}
\frac{dq}{dt}  &  =p\label{M-P1}\\
\frac{dp}{dt}  &  =-\omega^{2}(t)q+\frac{c}{q^{3}}\nonumber
\end{align}
where $c>0$ and $\omega(t+2\pi)=\omega(t).$

The system (\ref{M-P1}) is a 2$\pi-$periodic Hamiltonian system with
Hamiltonian function%
\begin{equation}
H(t,q,p)=\frac{1}{2}p^{2}+\frac{1}{2}(\omega^{2}(t)q^{2}+\frac{c}{q^{2}})
\label{M-PHamFct}%
\end{equation}

The Hamiltonian vector field%
\begin{equation}
X=p\frac{\partial}{\partial q}-(\omega^{2}(t)q-\frac{c}{q^{3}})\frac{\partial
}{\partial p} \label{M-PHvct}%
\end{equation}
can be written in the form%
\[
X=-\alpha(t)X_{2}-\beta(t)X_{3}%
\]
where $\alpha(t)=1-\omega^{2}(t)$ and $\beta(t)=\omega^{2}(t)+1.$ The vector
fields $X_{2},X_{3}$ are Hamiltonian vector fields%
\begin{align*}
X_{2}  &  =\frac{1}{2}(-p\frac{\partial}{\partial q}-(q+\frac{c}{q^{3}}%
)\frac{\partial}{\partial p})\\
X_{3}  &  =\frac{1}{2}(-p\frac{\partial}{\partial q}+(q-\frac{c}{q^{3}}%
)\frac{\partial}{\partial p})
\end{align*}
with Hamiltonian functions $H_{2},H_{3}$ given respectively by%
\begin{align*}
H_{2}(q,p)  &  =-\frac{1}{4}p^{2}+\frac{1}{4}(q^{2}-\frac{c}{q^{2}})\\
H_{3}(q,p)  &  =-\frac{1}{4}p^{2}-\frac{1}{4}(q^{2}+\frac{c}{q^{2}})
\end{align*}

Taking the Lie bracket between $X_{2}$ and $X_{3}$ we denote by $X_{1}$ the
Hamiltonian vector field given by%
\[
X_{1}=[X_{2},X_{3}]=\frac{1}{2}(q\frac{\partial}{\partial q}-p\frac{\partial
}{\partial p})
\]
with Hamiltonian function%
\[
H_{1}(q,p)=\frac{1}{2}pq
\]

$\bigskip$The commutation relations%
\begin{equation}
\lbrack X_{1},X_{2}]=-X_{3}\text{ },\text{ }[X_{2},X_{3}]=X_{1}\text{ },\text{
}[X_{3},X_{1}]=X_{2} \label{ConRel}%
\end{equation}
correspond to those of the Lie algebra $sp(1,%
\mathbb{R}
).$Then, the system (\ref{M-P1}) is a periodic Hamiltonian systems of Lie
type. Moreover, the relation $\left\{  H_{X},H_{Y}\right\}  =H_{[X,Y]}$ holds
for any $X,Y\in sp(1,%
\mathbb{R}
).$

Consider now, the Euler system on $sp(1,%
\mathbb{R}
)$ given by%
\begin{equation}
\frac{d\xi}{dt}=[\alpha(t)e_{2}+\beta(t)e_{3},\xi],\text{ \ }\xi=(\xi_{1}%
,\xi_{2},\xi_{3})\in%
\mathbb{R}
^{3} \label{EulEq}%
\end{equation}
where $e_{1},e_{2},e_{3}$ is the basis for $sp(1,%
\mathbb{R}
)$ with commuting relations given by (\ref{ConRel}). The system (\ref{EulEq})
takes the form
\begin{align*}
\frac{d\xi_{1}}{dt}  &  =-\beta(t)\xi_{2}+\alpha(t)\xi_{3}\\
\frac{d\xi_{2}}{dt}  &  =\beta(t)\xi_{1}\\
\frac{d\xi_{3}}{dt}  &  =\alpha(t)\xi_{1}%
\end{align*}
and can be written using the cross product in $%
\mathbb{R}
^{3}$%
\begin{equation}
-G\frac{d\xi}{dt}=\mu(t)\times\xi,\text{ \ }\xi\in%
\mathbb{R}
^{3} \label{EuEq2}%
\end{equation}
where $\mu(t)=(0,\alpha(t),\beta(t))$ and $G=diag(1,1,-1).$Note that
$K(\xi)=\xi_{1}^{2}+\xi_{2}^{2}-\xi_{3}^{2}$ is a first integral for
(\ref{EuEq2}). Moreover, taking into account that the periodic Euler system
(\ref{EuEq2}) possesses always periodic solutions \cite{[FLV]}, we have for
the 2$\pi-$periodic Milne-Pinney equation a 2$\pi-$periodic first integral of
the form
\[
I(t,q,p)=\ \frac{1}{2}pq\xi_{1}(t)+(q^{2}-p^{2}-\frac{c}{q^{2}})\xi
_{2}(t)-(q^{2}+p^{2}+\frac{c}{q^{2}})\xi_{3}(t)
\]
where $\xi(t)=(\xi_{1}(t),\xi_{2}(t),\xi_{3}(t))$ is a 2$\pi-$periodic
solution for the Euler linear system (\ref{EuEq2}).

\end{document}